\documentclass[journal,twoside,web]{ieeecolor}
\usepackage{tmi}
\usepackage{cite}
\usepackage{multirow}
\usepackage{amsmath,amssymb,amsfonts}
\usepackage{algorithmic}
\usepackage{graphicx}
\usepackage[referable]{threeparttablex}
\usepackage{textcomp}
\def\BibTeX{{\rm B\kern-.05em{\sc i\kern-.025em b}\kern-.08em
    T\kern-.1667em\lower.7ex\hbox{E}\kern-.125emX}}
\begin{document}
\title{SurgPointTransformer: Vertebrae Shape Completion with RGB-D Data}
\author{Aidana Massalimova, Florentin Liebmann, Sascha Jecklin, Fabio Carrillo,  Mazda Farshad and Philipp Fürnstahl
\thanks{Aidana Massalimova, Florentin Liebmann, Sascha Jecklin, Fabio Carrillo, and Philipp Fürnstahl are with Research in Orthopedic Computer Science, University Hospital Balgrist, University of Zürich, Zürich, 8008, Switzerland (corresponding author: aidana[dot]massalimova[at]balgrist[dot]ch).}
\thanks{ Mazda Farshad is with the Department of Orthopaedics, Balgrist University Hospital, University of Zurich, Zurich, 8008, Switzerland.}}

\maketitle

\begin{abstract}
State-of-the-art computer- and robot-assisted surgery systems heavily depend on intraoperative imaging technologies such as CT and fluoroscopy to generate detailed 3D visualization of the patient's anatomy. While imaging techniques are highly accurate, they are based on ionizing radiation and expose patients and clinicians. This study introduces an alternative, radiation-free approach for reconstructing the 3D spine anatomy using RGB-D data. Drawing inspiration from the 3D "mental map" that surgeons form during surgeries, we introduce SurgPointTransformer, a shape completion approach for surgical applications that can accurately reconstruct the unexposed spine regions from sparse observations of the exposed surface.

Our method involves two main steps: segmentation and shape completion. The segmentation step includes spinal column localization and segmentation, and vertebra-wise segmentation. The segmented vertebra point clouds are then subjected to SurgPointTransformer, which leverages an attention mechanism to learn patterns between visible surface features and the underlying anatomy. For evaluation, we utilize an ex-vivo dataset of nine specimens. Their CT data is used to establish ground truth data that were used to compare to the outputs of our methods. Our method significantly outperforms the state-of-the-art baselines, achieving an average Chamfer Distance of 5.39, an F-Score of 0.85, an Earth Mover's Distance of 0.011, and a Signal-to-Noise Ratio of 22.90 dB.

This study demonstrates the potential of our reconstruction method for 3D vertebral shape completion without ionizing radiation or invasive imaging. Our work contributes to computer-aided and robot-assisted surgery, advancing the perception and intelligence of these systems.

\end{abstract}


\section{Introduction}
Accurate surgical execution in spine surgery is essential due to the spine’s complex anatomy and the proximity of critical structures like the spinal cord, nerves, and major blood vessels. Precision is crucial for minimizing complications and improving patient outcomes. Traditionally,  surgeries are performed free-hand, where surgeons rely on their deep understanding of anatomy to avoid damaging vital structures. However, advancements in intraoperative navigation technologies, such as computer-assisted surgery (CAS) and surgical robotics, have been gradually integrated into the surgical pipeline  \cite{peng2020accuracy}. These technologies help localize the area of interest, verify fracture patterns, and ensure and the proper execution of the preoperative plan \cite{sielatycki2022state}. Fundamental components of these systems are intraoperative imaging technologies, allowing surgeons to monitor anatomical structures and surgical tools, visualizing unexposed or partially exposed areas like the pedicle region of a vertebra. As a result, procedures are more accurately aligned with preoperative plans.

Intraoperative fluoroscopy remains the most commonly used real-time imaging technology in spine surgery. Its mobile version, C-arm, can capture images from various angles in the horizontal plane with an angular range of 180°, allowing surgeons to visualize the spine from multiple perspectives. However, C-arm imaging is limited to two-dimensional imaging. O-arm, cone-beam CT (CB-CT), and intraoperative CT (iCT) overcome this limitation by offering three-dimensional (3D) imaging capabilities, providing real-time, comprehensive views of the surgical site. A study involving 107 spine surgery patients using the O-arm system reported an average patient radiation dose of 5.15 mSv (ranging from 1.48 to 7.64 mSv) \cite{costa2016radiation}.  Radiation exposure from computed tomography (CT) has been reported to range between 5.5 and 7.4 mSv per patient \cite{kendlbacher2022workflow}. For second-generation CB-CT, radiation exposure is approximately 19 mSv reported per spinal navigation procedure \cite{kendlbacher2022workflow}.

Non-radiative imaging techniques like ultrasound (US), multi-view stereo, time-of-flight cameras, and laser scanning are gaining interest in surgical applications due to their ability to provide real-time imaging and guidance without ionizing radiation. Several studies \cite{li2023robot,ji2015patient} have proposed landmark-based registration to align these methods with preoperative data. However, this approach presents challenges like a steep learning curve, longer operation times, and user frustration. To address these shortcomings, image- and surface-based registration methods have been proposed as a more accurate, automated, and faster solution \cite{faraji2020machine, liebmann2024automatic}. 

The growing capability of 3D optical reconstruction has fueled the field of 3D shape completion, a process that reconstructs a complete point cloud from partial observations. Accurate shape completion of RGB-D data still poses significant challenges due to sparse and irregularly distributed data, noise, and the need to preserve fine details. The two best-performing methods for 3D shape completion are VRCNet and PoinTr \cite{fei2022comprehensive}. VRCNet, a variational autoencoder-based method, models uncertainty in shape completion by generating multiple plausible completions. While these characteristics make VRCNet adaptable to scenarios with significant data loss, it often leads to over-smoothing and the loss of fine surface features \cite{pan2021variational}. PoinTr, a transformer-based method, excels at capturing long-range dependencies in 3D data \cite{yu2021pointr}. It uses transformers to model relationships between observed points and missing structures, effectively reconstructing complex and irregular shapes. AdaPoinTr, an extension of PoinTr, incorporates noise-reduction capabilities to tackle the common challenge of noisy real-world point clouds effectively \cite{adapointr}.

Only a few studies explored the effectiveness of shape completion approaches in the medical domain. Li et al. \cite{li2023anatomy} developed the Anatomy Completor framework, using a denoising autoencoder (DAE) to reconstruct anatomical shapes from incomplete CT images. Beetz et al. \cite{beetz2023multi} used the Point Completion Network (PCN) to reconstruct cardiac anatomy from 2D MRI slices. Similarly, Gafencu et al. \cite{gafencu2024shape} investigated shape completion for spinal anatomy using VRCNet from 3D US reconstructions. To the best of our knowledge, shape completion performance on RGB-D data of human anatomy has not yet been explored.

Our work addresses this knowledge gap, providing new insights and advancements in shape completion of the human anatomy. This work is inspired by the ability of experienced surgeons to mentally reconstruct unseen anatomical structures by complementing visible anatomy with their knowledge of human anatomy. We hypothesize that the attention mechanism of transformer networks can emulate this behavior by learning patterns and correlations between visible and hidden anatomical structures. Moreover, despite its significant dependence on upstream tasks like segmentation, most state-of-the-art methods treat shape completion as an isolated problem. We propose a fully-fletched pipeline for 3D reconstruction of the complete shape of the spine anatomy directly from raw RGB-D camera data of surgical procedures. Our method involves spinal column localization, spinal column segmentation, vertebra-level segmentation, and shape completion. We evaluated our method on the SpineDepth dataset \cite{liebmann2021spinedepth} and using it as a baseline, and benchmarked it against Gafencu et al.’s \cite{gafencu2024shape} work. Our approach eliminates the need for traditional registration processes, intraoperative radiation exposure, and addresses key concerns in current surgical practices.

\section{Methods}
\begin{figure*}[ht]
    \centering
    \includegraphics[width=\linewidth]{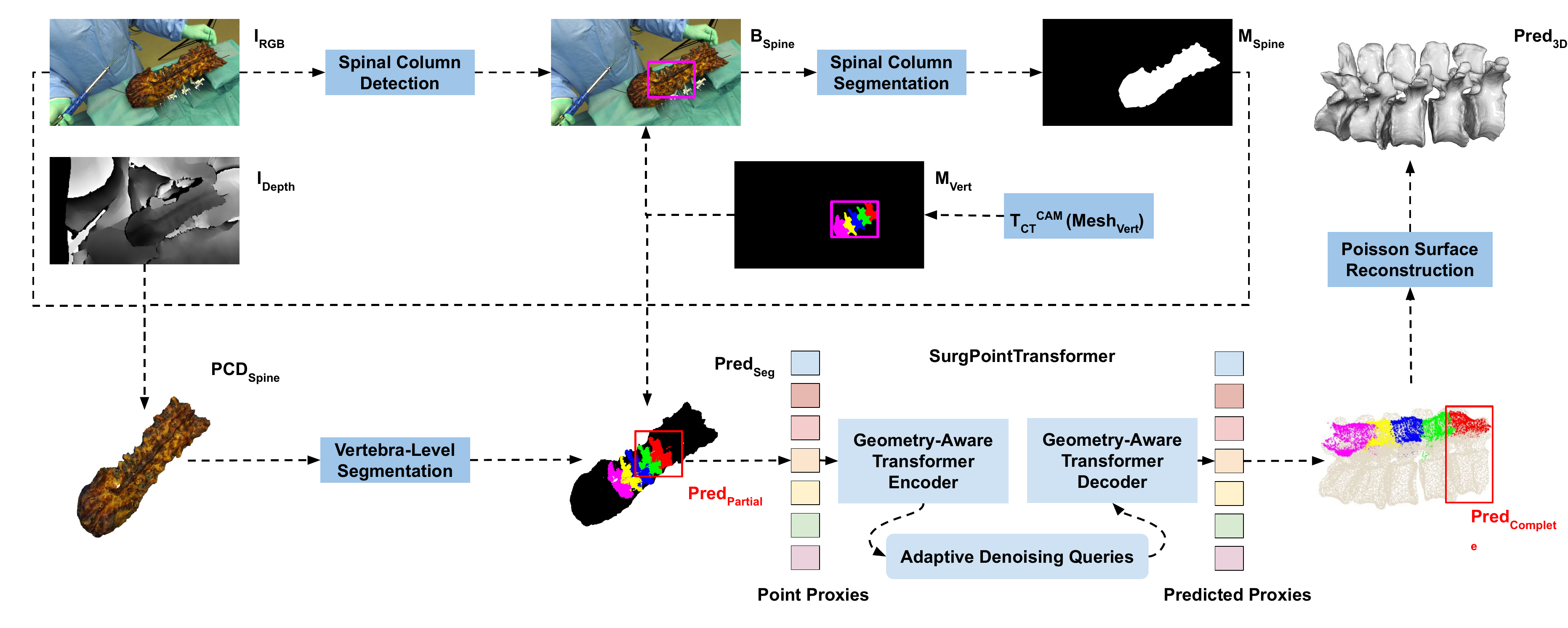}
   \caption{Pipeline of the proposed method for vertebrae shape completion from RGB-D data: $I_{RGB}$ from the RGB-D data is fed into the spinal column detector, which localizes the spine's position with a bounding box $B_{Spine}$ (indicated in fuchsia). $B_{Spine}$ and $I_{RGB}$ are then passed to the spine segmentation model, resulting in a spinal segmentation mask ($M_{Spine}$).This mask then applied to  $I_{RGB}$ and  $I_{Depth}$ to produce  $PCD_{Spine}$. The vertebra-level segmentation module produces color-coded segmentation ($Pred_{Seg}$) for each vertebra level, where red, green, blue, yellow, fuchsia, and black colors correspond to  L1, L2, L3, L4, L5, and background, respectively. The segmented vertebra point clouds ($Pred_{Partial}$) are input to our SurgPointTransformer, which reconstructs the complete shape of each vertebra ($Pred_{Complete}$). The completed point clouds are converted into 3D meshes ($Pred_{3D}$).}
    \label{fig:pipeline}
\end{figure*}

Our pipeline, built around shape completion, comprises spinal column localization, spinal column segmentation, and vertebra-level segmentation as preprocessing steps (see Figure \ref{fig:pipeline}). First, we capture color ($I_{RGB}$) and depth data ($I_{Depth}$), as described in Section \ref{sec:rgbd dataset}. Our method localizes the spinal column with a bounding box ($B_{Spine}$) and segments the spinal column by generating a spine segmentation mask ($M_{Spine}$) and then applies $M_{Spine}$ on the RGB-D data to produce spinal point cloud ($PCD_{Spine}$). It is followed by a vertebra-wise segmentation resulting in color-coded point cloud ($Pred_{Seg}$), detailed in Section \ref{sec:segmentation}. The resulting vertebra segments ($Pred_{Partial}$) are then input to our shape completion network SurgPointTransformer (Section \ref{sec: shape prediction}). The predicted point clouds ($Pred_{Complete}$) are then converted to 3D meshes ($Pred_{3D}$) using Poisson surface reconstruction. We detail the evaluation metrics for assessing the method's effectiveness in Section \ref{sec:evaluation}. All training and evaluations were performed using an NVIDIA Tesla V100 GPU with 16GB RAM and Python 3.11.5 and PyTorch 2.4.0.

\subsection{Dataset Description}
\label{sec:rgbd dataset}
At our institution, ten ex-vivo spine surgeries were conducted by a surgeon in a realistic surgical environment. We used two ZED mini stereo cameras (Stereolabs Inc., San Francisco, CA, USA) to simultaneously acquire the recordings from two views, with pose annotations provided for individual spinal vertebrae. Each specimen underwent 40 recording sessions with varying camera viewpoints. Additionally, the dataset included 3D meshes of L1-L5 vertebrae extracted from preoperative CT data using commercial medical imaging software (Mimics Medical, Materialise NV, Leuven, Belgium). This dataset was later made public and provided to the community as described in \cite{liebmann2021spinedepth}.  To maintain consistency with prior research, the first specimen, whose surgical exposure significantly differed from the others, was excluded.

We randomly selected surface reconstruction from 160 frames of different steps in the surgery and two perspectives, obtaining a dataset including 320 frames of RGB-D data per specimen (160 frames x 2 perspectives) paired with ground truth 3D vertebra meshes. We adopted a leave-one-specimen-out cross-validation approach on each dataset. Consequently, the dataset comprised nine folds, each containing 2'560 samples in the training set and 320 samples in the validation set. This approach allowed us to systematically evaluate the performance of our shape completion pipeline while ensuring robustness and generalizability across different specimens.

\subsection{Segmentation}
\label{sec:segmentation}
The spine segmentation method involves the following three stages:

\textbf{Spinal Column Localization}: We trained a YOLOv8 network \cite{yolov8_ultralytics} to detect the spine region in color images. YOLOv8 outputs a bounding box ($B_{Spine}$) around the detected spine region. The ground truth data for training were generated automatically from 3D vertebra meshes ($Mesh_{Vert}$). We applied a respective transformation matrix ($T_{CT}^{CAM}$), available through the calibration chain described in \cite{mendelsohn2016patient}, to convert $Mesh_{Vert}$ from CT to camera space to align them with the RGB-D data. Using the camera's intrinsic parameters, we rendered a vertebra-level mask ($M_{Vert}$) of the lumbar spine from the 3D meshes. The bounding box of $M_{Vert}$ was used as ground truth for training YOLOv8. The model was trained separately for each fold, with training limited to three epochs. 

\textbf{Spinal Column Segmentation}: This step refines the region-of-interest of spinal column to lower computational costs. Segment Anything Model's (SAM) \cite{kirillov2023segany} ability to adapt to different input types and its robustness across various image domains make it a crucial component in this segmentation pipeline. The bounding box  ($B_{Spine}$) and the color image  ($I_{RGB}$) are fed into SAM to generate a segmentation mask of the spine ($M_{Spine}$). $M_{Spine}$ was further applied on $I_{Depth}$ and $I_{RGB}$ to generate spinal point cloud ($PCD_{Spine}$).

\textbf{Vertebra-Level Segmentation}: We employed the PointNet++ network \cite{qi2017pointnet++} to segment the spinal column point cloud into individual vertebra levels. PointNet++ builds upon the original PointNet architecture \cite{qi2017pointnet} by addressing its limitations in capturing local geometric structures. It employs a hierarchical approach to feature learning through abstraction layers. 

Each point in $PCD_{Spine}$ was labeled with different vertebral levels, including L1 to L5 and background, to train the multi-class point cloud segmentation network. We applied $M_{Vert}$ on $PCD_{Spine}$ to generate ground truth segmentation labels ($GT_{Seg}$). We downsampled $PCD_{Spine}$ to 10'000 points using open3d 0.18.0 \cite{open3d} library to reduce computational complexity during training. However, the full-resolution, dense point cloud was used during inference.

We explored two input configurations for PointNet++: one using only the Cartesian coordinates (XYZ) of the points and another using both Cartesian coordinates and RGB color information (XYZRGB). The model was trained nine times per experiment with a batch size 16. The training was performed using the Adam optimizer, starting with a learning rate of 1e-3 and betas set to (0.9 and 0.999) to control momentum and variance adaptation, respectively. We used Negative Log-Likelihood (NLL) loss, which is well-suited for our multi-class classification task.  We applied a learning rate scheduler (StepLR) to adjust the learning rate every 20 epochs, reducing it by half (gamma=0.5) to fine-tune the training process and improve convergence.

\subsection{Shape Completion}
\label{sec: shape prediction}
SurgPointTransformer utilizes AdaPoinTr \cite{adapointr}, which features an encoder-dethe coder structure with geometry-aware blocks designed to capture and model explicit geometric relationships. We designed SurgPointTransformer to predict spine anatomy with an encoder depth of 6 layers and a decoder depth of 8 layers. We used a Dynamic Graph Convolutional Neural Network (DGCNN) \cite{wang2019dynamic} as the feature extractor, setting the number of nearest neighbors (kNN) to 6 and 8 for various operations. Each transformer block incorporates multi-head attention with six attention heads and 384 hidden dimensions. Through an adaptive query mechanism, this helps manage significant noise resulting from RGB-D sensor inaccuracies and self-occlusions.

Our network processed the point cloud of each vertebra ($GT_{Partial}$), which was obtained from $GT_{Seg}$. Each fold consisted of  12'800 (2'560 samples x 5 vertebrae) samples in the training set and 1600 (320 samples x 5 vertebrae) in the validation set. We downsampled $GT_{Partial}$ during training to 2048 points per vertebra as the input data for our model. Our ground truth ($GT_{Complete}$) consisted of 3D meshes derived from CT scans, which were also downsampled to 4'098 points. We trained the model nine times with the AdamW optimizer, a batch size of 32, an initial learning rate of 1e-4, and a weight decay 5e-4. We used Chamfer's distance (CD) loss as the evaluation metric to assess the model's performance. 

We implemented VRCNet from \cite{gafencu2024shape} as a benchmark method and trained it on our data, using the same loss function as in the original work, which combines Kullback–Leibler (KL) divergence loss and CD loss. The model was trained separately for each fold with a batch size of 32. We used the Adam optimizer with a starting learning rate 1e-4 and betas set to (0.9, 0.999). A learning rate scheduler (StepLR) was utilized to reduce the learning rate by 0.7 every 40 epochs to improve training stability.

\subsection{Evaluation}
\label{sec:evaluation}
To evaluate the vertebra-level segmentation module, we used the following metrics:

\textbf{Accuracy}: The proportion of correctly classified points in $Pred_{Seg}$  compared to $GT_{Seg}$, indicating the overall correctness of the segmentation.

\textbf{Intersection over Union (IoU)} evaluates the accuracy of segmentation models by measuring the overlap between $Pred_{Seg}$ and $GT_{Seg}$ in point clouds.

For the shape completion task, we assessed performance using:

\textbf{Chamfer Distance (CD)} measures the average distance between corresponding points in $Pred_{Complete}$ and $GT_{Complete}$. Lower CD values indicate better alignment between $Pred_{Complete}$ and $GT_{Complete}$.
\begin{equation}
\begin{split}
\text{CD} = \frac{1}{|Pred_{\text{Complete}}|} \sum_{p \in Pred_{\text{Complete}}} \min_{g \in GT_{\text{Complete}}} \| p - g \|_2^2 \\
+ \frac{1}{|GT_{\text{Complete}}|} \sum_{g \in GT_{\text{Complete}}} \min_{p \in Pred_{\text{Complete}}} \| g - p \|_2^2
\end{split}
\end{equation}

\textbf{CD of the exposed, visible surface (CD\_top)} measures the average distance between points in the predicted ($Pred_{Top}$) and ground truth ($GT_{Top}$) point clouds, specifically for the visible surface. This refers to the area that overlaps with the input point cloud, as determined by the segmentation module.

\textbf{CD of the unexposed, invisible surface (CD\_bottom)} measures the average distance between points in the predicted ($Pred_{Bottom}$) and ground truth ($GT_{Bottom}$) point clouds for the invisible surface. This refers to the areas that do not overlap with the input point cloud and are instead predicted by the model.

\textbf{F-Score} combines precision and recall into a single metric by calculating their harmonic mean. It balances the trade-off between false positives and false negatives in the predictions. We used the F1-score@1\% proposed by \cite{tatarchenko2019single}.

\textbf{Intersection over Union of the input point cloud (IoU\_input)}: To assess how variations in the input point cloud affect results, we calculated IoU between $Pred_{Partial}$ and $GT_{Complete}$.

\textbf{Earth Mover’s Distance (EMD)}: Also known as Wasserstein distance, EMD measures the minimum cost of transforming the $Pred_{Complete}$ into $GT_{Complete}$, considering the distance between corresponding points. Lower EMD values indicate better shape alignment.
\begin{equation}
\centering
\text{EMD} = \min_{\phi: Pred_{Complete} \to GT_{Complete}} \sum_{p \in Pred_{Complete}} \| p - \phi(p) \|_2
\end{equation}
, where $\phi$ is a bijective mapping from points in $Pred_{Complete}$ to $GT_{Complete}$.

\textbf{Signal-to-Noise Ratio (SNR)} measures the level of a desired signal to the background noise level.
The SNR is estimated according to the following equations \cite{zeng20193d}:
\begin{equation}
P_{\text{signal}} = \frac{1}{N} \sum_{g \in GT_{Complete}} \| g - \mu \|^2
\end{equation}
, where $N$ and $\mu$ are the number of points in  and centroid of  $GT_{Complete}$, respectively.
\begin{equation}
P_{\text{noise}} = \frac{1}{N} \sum_{g \in GT_{Complete},p \in Pred_{Complete}} \| g - p \|^2
\end{equation}
\begin{equation}
\text{SNR (dB)} = 10 \log_{10}\frac{P_{\text{signal}}}{P_{\text{noise}}} 
\end{equation}
Higher SNR means the signal is much stronger than the noise, resulting in clearer or higher-quality information.

We also examined the correlations between the data characteristics (specimen, vertebra level) and localization and segmentation performance on the overall effectiveness of our pipeline, using the Pearson correlation method to assess the strength and direction of the relationships.

\section{Results}

\begin{figure*}
    \centering
    \includegraphics[width=\linewidth]{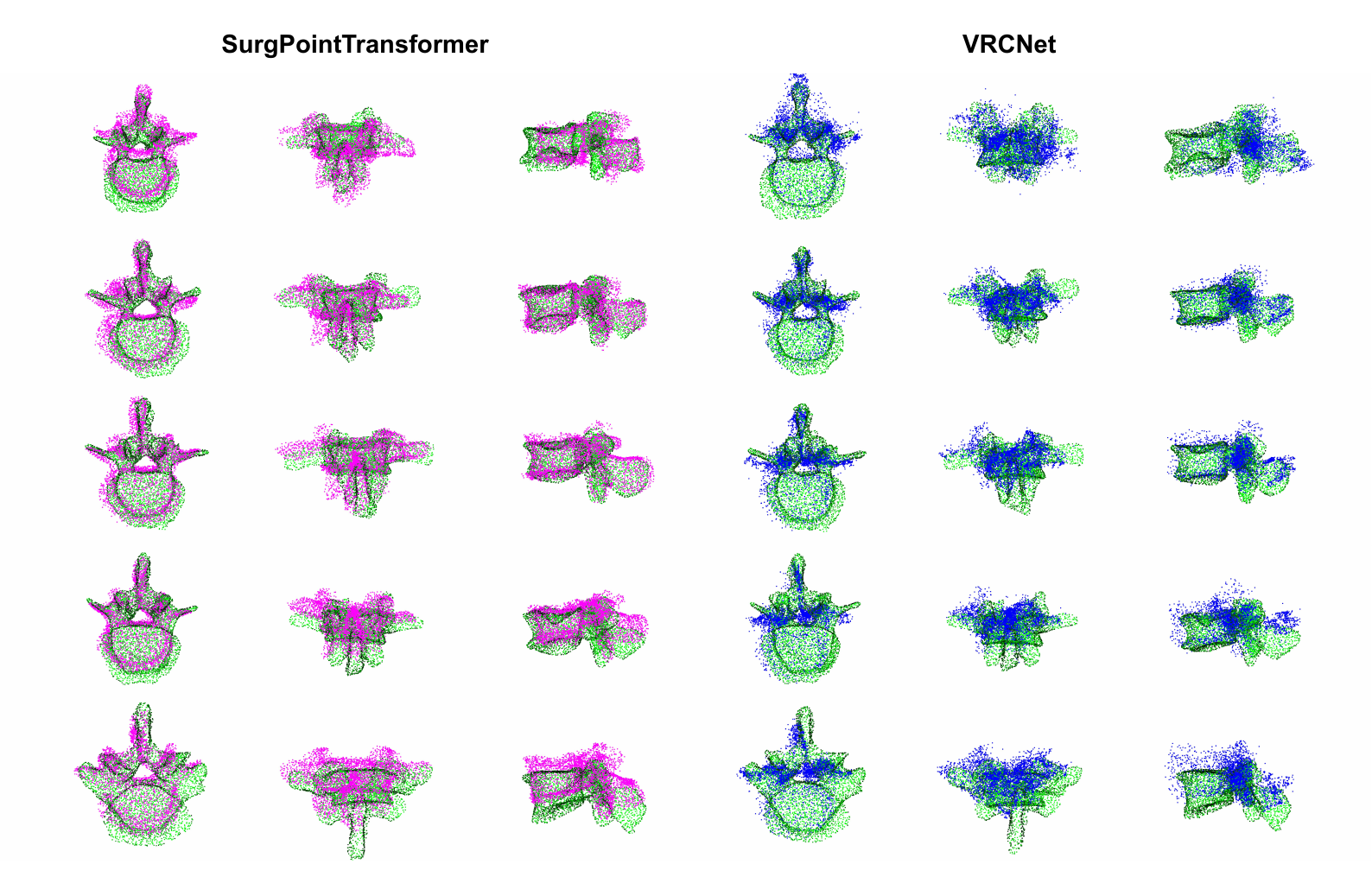}
    \caption{Visual Representation of Segmentation and Shape Completion Outputs for L1-L5 Vertebrae from VRCNet and SurgPointTransformer. This figure shows axial, coronal, and sagittal views of shape completion outputs for the L1 (first row) through L5 vertebrae (last row). The outputs from our approach are in fuchsia, and from the state-of-the-art baseline, VRCNet, are in blue. Both are overlaid with the ground truth point cloud in green. The figure also includes evaluation scores for the shape completion results.}
    \label{fig:visual results}
\end{figure*}

\begin{figure*}
    \centering
    \includegraphics[width=0.8\linewidth]{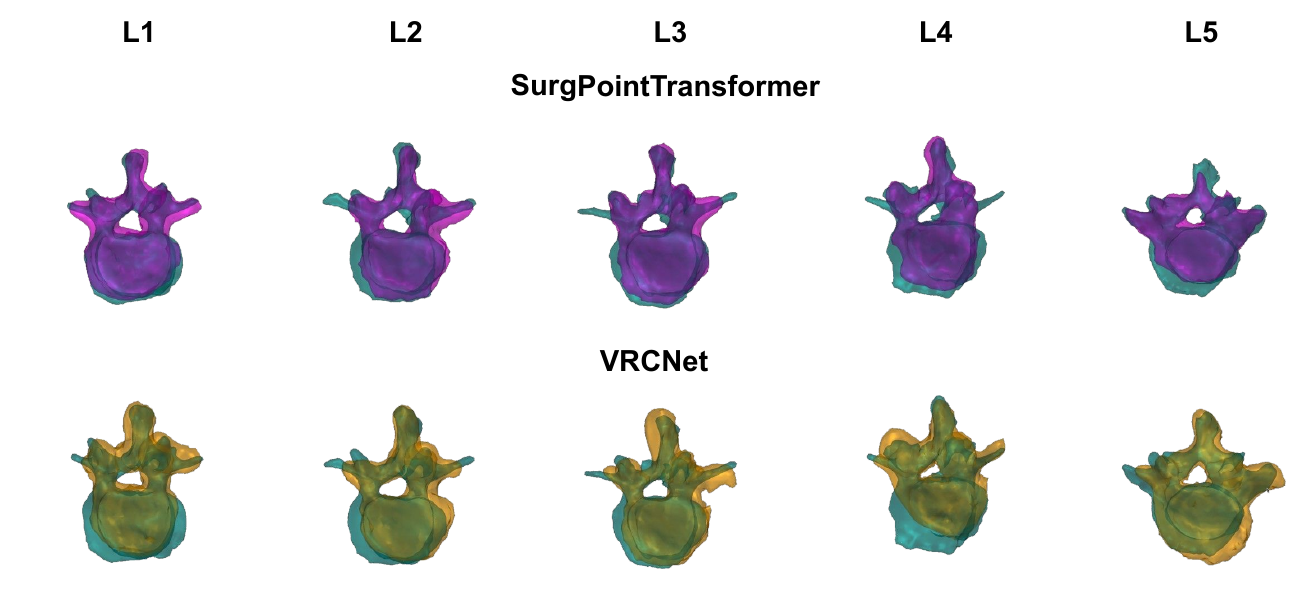}
    \caption{ Poisson surface reconstruction \cite{kazhdan2006poisson} applied on SurgPointTransformer (shown in fuchsia) and VRCNet (shown in blue) outputs overlaid on the ground truth 3D meshes (shown in green) in axial and lateral views. }
    \label{fig:3D surface reconstruction}
\end{figure*}

The proposed shape completion method results are demonstrated in Table \ref{Shape Completion Results per vertebra}. Our SurgPointTransformer achieved an average CD of 5.39, indicating excellent performance in shape completion. The method also showed a high F-Score of 0.85, reflecting its ability to produce well-defined and accurate shapes. The EMD value was notably low at 0.011, suggesting that the completed shapes closely aligned with the ground truth. Furthermore, SurgPointTransformer delivered a high SNR of 22.90 dB, emphasizing the clarity and reduced noise in the completed shapes. Visual inspections further support these findings, as SurgPointTransformer's outputs are characterized by smoother surfaces and fewer artifacts. This is particularly evident in Figure \ref{fig:visual results} and Figure \ref{fig:3D surface reconstruction}, where the results from SurgPointTransformer, shown in fuchsia, display more uniform and less noisy shapes. The results in Table \ref{Overall our results per specimen} show that Specimen 2 had the best overall performance, achieving an accuracy of 0.74, an IoU of 0.60, a CD of 4.10, an F-Score of 0.94, an EMD of 0.008, a CD\_top of 3.90, a CD\_bottom of 4.43, and an SNR of 23.88 dB. Its high IoU\_input value of 0.37 could explain this strong performance. In contrast, Specimen 8 demonstrated the poorest performance, with the lowest IoU\_input of 0.20, which may have contributed to its weaker results across all metrics. 
\begin{table*}
\centering
\caption{Shape Completion Results for our approach against VRCNet. This table shows the performance metrics for shape completion, including IoU, CD, F-Score, EMD, CD\_top, CD\_bottom, and SNR. Results are provided for each vertebra level and their averages. The superior results between the two methods are highlighted in bold. $\mu$ means the average value over all vertebrae levels.}
\begin{tabular}{rrcccccc}
\textbf{Method}                     & \textbf{Class}   & \textbf{CD}   & \textbf{F1}   & \textbf{EMD}   & \textbf{CD\_top} & \textbf{CD\_bottom} & \textbf{SNR}   \\
\hline
\hline
\multirow{6}{*}{\textbf{SurgPointTransformer}} & 1                 & \textbf{5.58} & 0.82          & \textbf{0.012} & \textbf{5.48}    & \textbf{6.37}       & \textbf{23.95} \\
                                    & 2                       & \textbf{5.59} & 0.82          & \textbf{0.011} & \textbf{5.36}    & \textbf{6.42}       & \textbf{23.37} \\
                                    & 3              & \textbf{5.25} & 0.85          & \textbf{0.011} & \textbf{5.23}    & \textbf{5.83}       & 22.16          \\
                                    & 4                       & \textbf{5.15} & 0.85          & \textbf{0.011} & \textbf{5.13}    & \textbf{5.78}       & 22.75          \\
                                    & 5                     & \textbf{5.39} & 0.84          & \textbf{0.011} & \textbf{5.24}    & \textbf{5.95}       & 22.32          \\
                                    & Average        & \textbf{5.39} & 0.85          & \textbf{0.011} & \textbf{5.10}    & \textbf{5.86}       & \textbf{22.90} \\
\multirow{6}{*}{\textbf{VRCNet}}    & 1              & 6.31          & \textbf{0.86} & 0.020          & 6.26             & 6.40                & 23.46          \\
                                    & 2              & 6.27          & \textbf{0.85} & 0.021          & 6.14             & 6.49                & 23.02          \\
                                    & 3                      & 6.21          & \textbf{0.86} & 0.020          & 6.22             & 6.29                & \textbf{22.52} \\
                                    & 4                        & 6.02          & \textbf{0.87} & 0.020          & 5.99             & 6.10                & \textbf{22.86} \\
                                    & 5                    & 6.05          & \textbf{0.87} & 0.020          & 5.98             & 6.22                & \textbf{22.63} \\
                                    & Average             & 6.17          & \textbf{0.86} & 0.020          & 6.12             & 6.30                & 22.89         
\end{tabular}
\label{Shape Completion Results per vertebra}

\end{table*}

Compared to the state-of-the-art based on VRCNet, SurgPointTransformer achieved a significantly more accurate reconstruction with respect to CD and EMD. The visual outputs from VRCNet, depicted in blue in Figure \ref{fig:visual results}, show more noticeable noise and less uniformity than SurgPointTransformer’s results. The average CD for VRCNet was higher at 6.17. Although VRCNet achieved a slightly better F-Score of 0.86, this advantage is offset by its higher EMD of 0.020, which reflects more significant discrepancies between the predicted and actual shapes. The point clouds from VRCNet were resampled first before applying Poisson reconstruction for visualization purposes. 

\begin{figure*}[ht]
    \centering
    \includegraphics[width=0.8\linewidth]{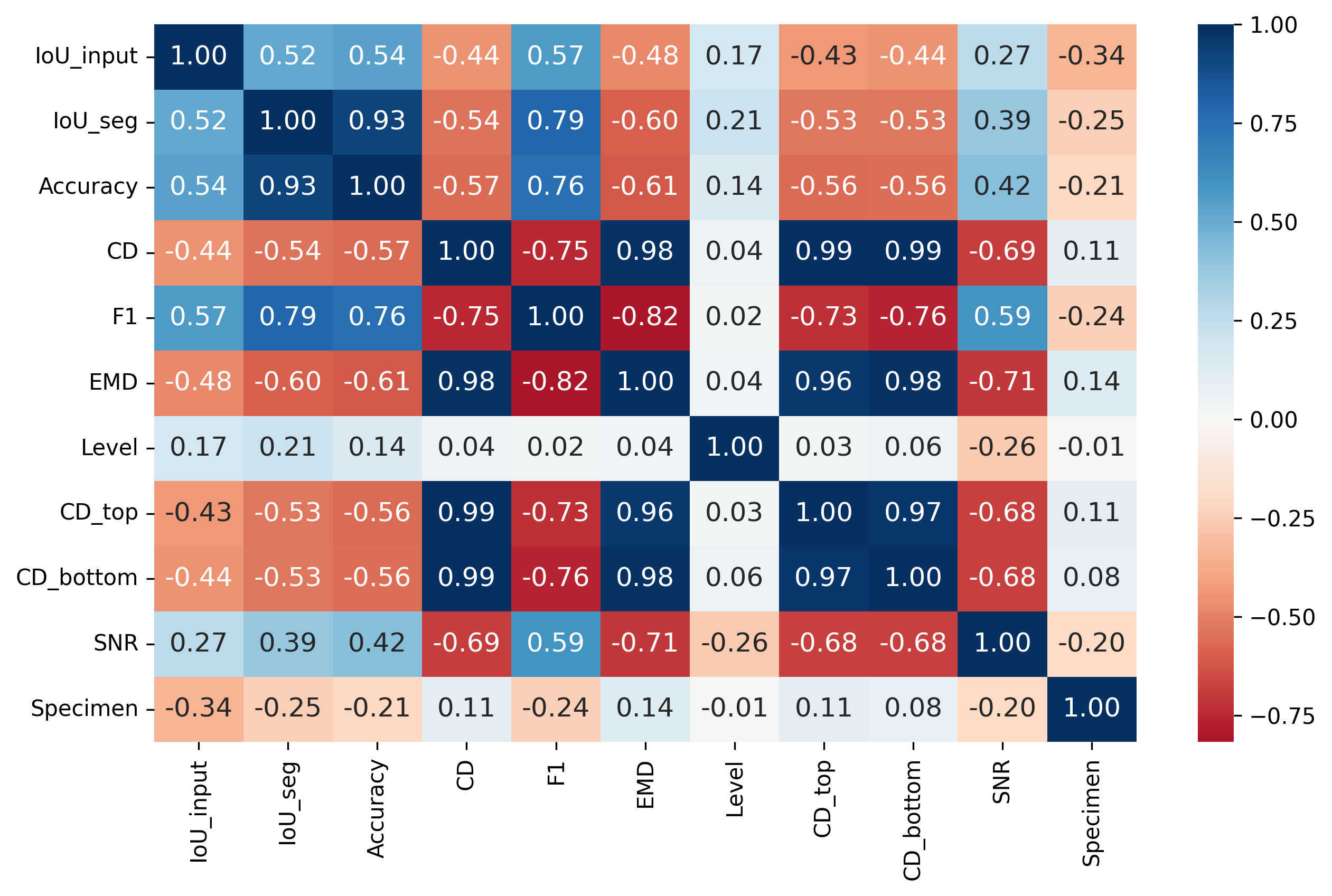}
    \caption{Correlation matrix between variables (specimen and vertebrae level) and evaluation matrices.}
    \label{fig:correlation matrix}
\end{figure*}

The Pearson correlation examination between data characteristics (specimen, vertebra level), segmentation, and shape completion performance are depicted in Figure \ref{fig:correlation matrix}. The results show that IoU\_input moderately positively correlated with IoU and accuracy from the segmentation results and the F-score from the shape completion tasks. It also showed a moderate negative correlation with Chamfer's Distance values (CD, CD\_top, and CD\_bottom), indicating that better input segmentation leads to lower errors in shape completion. Additionally, the SNR was found to have a strong negative correlation with CD, EMD, CD\_top, and CD\_bottom, while it showed a moderate positive correlation with the F-Score. This suggests higher SNR values are associated with better shape completion performance and lower error rates. Interestingly, specimen number and vertebra level did not correlate with any of the evaluated metrics, implying that these variables did not influence performance variations.

Regarding the performance of the localization and segmentation components, the results in Table \ref{Table:segmentation} show the XYZRGB input configuration consistently outperformed the XYZ configuration. The overall IoU increased from 0.69 with the XYZ input to 0.72 with the XYZRGB input, and accuracy improved from 0.79 to 0.83. Specifically, the L1 and L2 vertebrae showed significant improvements in IoU, increasing from 0.30 to 0.40 and from 0.36 to 0.45, respectively. These results highlight the advantage of using RGB information and XYZ data for more accurate segmentation.

\begin{table}[ht]
\centering
\caption{The vertebra-wise segmentation results: IoU and Accuracy for each class and overall, comparing two different input configurations. The best results comparing two input configurations are highlighted in bold.}

\begin{tabular}{rcccc}
\multicolumn{1}{l}{} & \multicolumn{2}{c}{\textbf{XYZ}} & \multicolumn{2}{c}{\textbf{XYZRGB}} \\
\hline \hline
\textbf{Class}       & \textbf{IoU} & \textbf{Accuracy} & \textbf{IoU}   & \textbf{Accuracy}  \\
\hline \hline

Overall    & 0.69         & 0.79              & \textbf{0.72}  & \textbf{0.83}      \\
L1         & 0.30         & 0.38              & \textbf{0.40}  & \textbf{0.54}      \\
L2         & 0.36         & 0.46              & \textbf{0.45}  & \textbf{0.60}      \\
L3         & 0.42         & 0.57              & \textbf{0.53}  & \textbf{0.71}      \\
L4         & 0.36         & 0.47              & \textbf{0.49}  & \textbf{0.61}      \\
L5         & 0.39         & 0.49              & \textbf{0.51}  & \textbf{0.62}      \\
Background & 0.94         & 0.93              & \textbf{0.96}  & \textbf{0.92}     
\end{tabular}
\label{Table:segmentation}
\end{table}

\begin{figure*}
    \centering
    \includegraphics[width=0.8\linewidth]{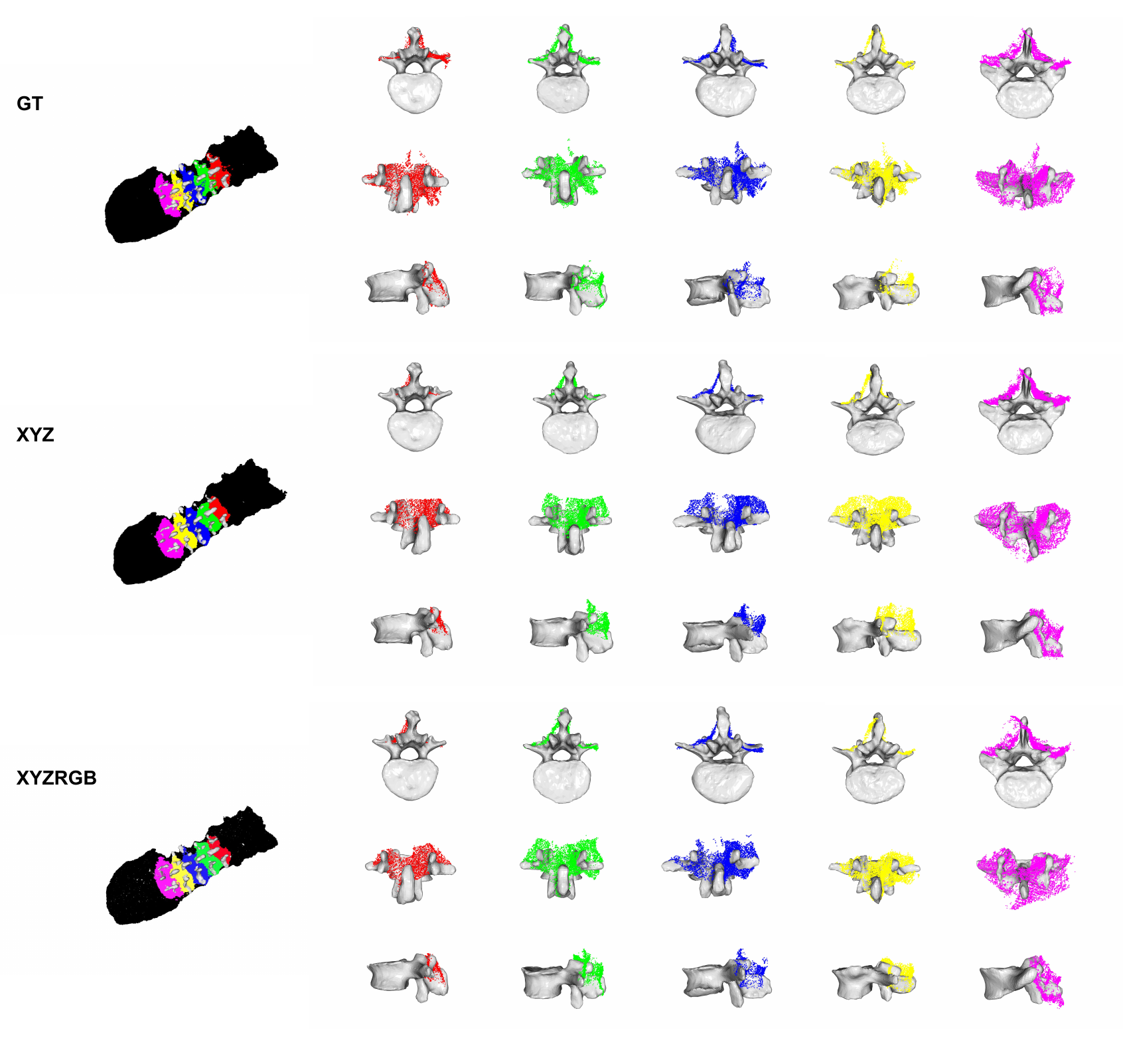}
    \caption{Segmentation results for XYZ and XYZRGB inputs are shown with ground truth. Point clouds are overlaid on 3D vertebra meshes, with L1 (red), L2 (green), L3 (blue), L4 (yellow), and L5 (pink) highlighted separately.}
    \label{fig:segmentation results visual}
\end{figure*}

\begin{table*}[ht]
\centering
\caption{Performance Metrics for Segmentation and Shape Completion Tasks Using Our Proposed Method. This table displays the results for each specimen, including IoU\_input, IoU\_seg, and Accuracy for the segmentation task, as well as CD, F-Score, EMD, CD\_top, CD\_bottom, and SNR for the shape completion task. The superior results for each specimen are highlighted in black bold, while the least are highlighted in blue bold.}

\begin{tabular}{rccccccccc}
\textbf{Specimen} & \textbf{IoU\_input}                   & \textbf{IoU} & \textbf{Accuracy}                    & \textbf{CD}                          & \textbf{F1}                          & \textbf{EMD}                          & \textbf{CD\_top}                     & \textbf{CD\_bottom}                  & \textbf{SNR}                          \\
\hline \hline
\textbf{2}        & \textbf{0.37}                         & \textbf{0.60}     & \textbf{0.74}                        & \textbf{4.10}                        & \textbf{0.94}                        & \textbf{0.008}                        & \textbf{3.90}                        & \textbf{4.43}                        & \textbf{23.88}                        \\
\textbf{3}        & 0.28                                  & 0.56              & 0.73                                 & 5.31                                 & 0.86                                 & 0.011                                 & 4.68                                 & 6.18                                 & 23.05                                 \\
\textbf{4}        & 0.27                                  & 0.56              & 0.68                                 & 5.57                                 & 0.83                                 & 0.011                                 & 4.60                                 & 7.02                                 & 23.80                                 \\
\textbf{5}        & 0.27                                  & 0.47              & 0.58                                 & 6.17                                 & 0.80                                 & 0.012                                 & 6.12                                 & 6.37                                 & 22.42                                 \\
\textbf{6}        & 0.26                                  & 0.52              & 0.72                                 & 5.23                                 & 0.85                                 & 0.011                                 & 4.97                                 & 5.63                                 & 22.12                                 \\
\textbf{7}        & 0.26                                  & 0.47              & {\textcolor{blue}{ \textbf{0.56}}} & 5.13                                 & 0.86                                 & 0.010                                 & 4.98                                 & 5.54                                 & {\textcolor{blue}{ \textbf{21.44}}} \\
\textbf{8}        & {\textcolor{blue}{ \textbf{0.20}}}  & {\textcolor{blue}{ \textbf{0.41}}} & 0.58                                 & {\textcolor{blue}{ \textbf{7.43}}} & {\textcolor{blue}{ \textbf{0.72}}} & {\textcolor{blue}{ \textbf{0.015}}} & {\textcolor{blue}{ \textbf{7.25}}} & {\textcolor{blue}{ \textbf{7.77}}} & 23.70                                 \\
\textbf{9}        & 0.25                                  & 0.49              & 0.71                                 & 5.87                                 & 0.83                                 & 0.011                                 & 5.53                                 & 6.24                                 & 23.09                                 \\
\textbf{10}       & 0.21                                  & 0.54              & 0.67                                 & 5.99                                 & 0.80                                 & 0.013                                 & 6.18                                 & 5.84                                 & 22.56                                 \\
\textbf{Average}  & 0.26                                  & 0.51              & 0.66                                 & 5.65                                 & 0.83                                 & 0.011                                 & 5.36                                 & 6.11                                 & 22.90                                
\end{tabular}
\label{Overall our results per specimen}
\end{table*}

\section{Discussion}

In this study, we presented the first approach utilizing RGB-D data for shape completion of spinal anatomy. Our shape completion method demonstrated superior performance in both accuracy and uniformity compared to the state-of-the-art method, VRCNet. SurgPointTransformer achieved an average CD of 5.39, an F-Score of 0.85, an EMD of 0.011, and an SNR of 22.90 dB. 

Other related work in the medical field has proposed shape completion for imaging data such as MRI, CT, and ultrasound. These modalities generally have a higher signal-to-noise ratio and fewer occlusions, presenting different challenges for shape completion. For instance, Beetz et al. \cite{beetz2023multi} reconstructed 3D cardiac structures from 2D cine MRI images and reported an average CD of 1.14 using a point cloud convolutional network (PCCN) on a large dataset of 1000 subjects. Although the input data is sparse, it still provides a more complete representation of the anatomy, as the 2D slices sample the entire structure.  Similarly, Li et al.’s Anatomy Completor \cite{li2023anatomy}, which reconstructed anatomical shapes from partial CT scans, achieved dice scores ranging from 0.865 to 0.931 using data from 737 subjects. While these scores are promising, their input partial CT scans present more information, specifically from 60\% to 90\% of the complete anatomy, whereas ours only include 26\%.

In contrast to medical imaging data, where the bone surface is clearly visible, one significant factor that influences the quality of the RGB-D input data and, consequently, the performance of shape completion - is the presence of soft tissue in the RGB-D scans, which complicates the segmentation process. This challenge is reflected in our average input IoU, which is only 0.26, indicating that a considerable portion of the vertebra remains hidden or partially occluded in the input data. This low IoU highlights the challenge of predicting accurate 3D shapes under these conditions, underscoring the effectiveness of our approach in achieving strong performance despite these difficulties. In contrast to statistical shape models (SSM) \cite{meng2020learning}, which rely on generalized templates and prior knowledge of anatomical variability, our approach leverages the transformer's capabilities to predict detailed and individualized reconstructions from incomplete and noisy input data. Despite the challenges posed by low IoU and occlusions, SurgPointTransformer’s more uniform and less noisy point cloud distributions enable it to produce higher-quality reconstructions than VRCNet, which tended to overfit the exposed regions of the input data (see Figure \ref{fig:visual results}).

Our results further suggest that uniformly distributed point clouds, especially with higher resolution, are essential for accurate shape completion. As demonstrated, the Poisson reconstruction meshing process results in higher-quality meshes, which is crucial for downstream tasks such as trajectory planning \cite{caprara2022bone}. SurgPointTransformer's more even point distribution contrasts with VRCNet’s concentration of points in localized, exposed regions, leading to noisier reconstructions. This finding underscores the importance of generating denser and more evenly distributed point clouds, particularly in regions not directly captured by the input.

Segmentation accuracy remains a critical factor influencing shape completion performance. Errors in the segmentation stage propagate throughout the pipeline, affecting the final output quality. Although SurgPointTransformer is robust enough to compensate for segmentation inaccuracy to a certain extent, improvements in this area would further enhance shape completion accuracy. Our vertebra-wise segmentation, which involved XYZ and XYZRGB configurations, demonstrated better results when using the XYZRGB configuration, incorporating both spatial and color information. This configuration achieved an IoU of 0.72 and an accuracy of 0.83. These results are comparable to the results from Ye et al. \cite{yanni2021real}, where they reported an average IoU of 0.72 and an accuracy of 0.89 in segmenting real-world noisy scene dataset using the Point Noise-Adaptive Learning (PNAL) framework. However, environmental factors such as lighting and occlusions from overlying tissues can negatively impact RGB-D data quality, making segmentation and subsequent shape completion challenging. Further research is needed to understand the effects of such conditions on segmentation and completion outcomes and to develop methods to mitigate their influence.

Our future work will focus on translating SurgPointTransformer into patient treatment. One critical step is integrating RGB-D cameras into the operating room environment, positioning them close to the surgical site to capture higher-quality real-time data. Another step will be validating our method on in-vivo data, including patients from multiple centers and surgeons, to ensure generalizability. Testing the approach on a broader range of subjects and anatomical variations will ensure its robustness and clinical reliability. Enhanced calibration between the RGB-D data and ground truth 3D meshes will also improve the system's accuracy. By addressing these challenges, we aim to pave the way for integrating our approach to surgical navigation systems that provide radiation-free 3D anatomical reconstructions in real-time.

\section{Conclusion}

This study confirms the hypothesis that RGB-D data can effectively predict the complete 3D shape of spinal vertebrae without radiation exposure. Our method significantly outperforms state-of-the-art baselines, achieving an average Chamfer Distance of 5.39, an F-Score of 0.85, an Earth Mover's Distance of 0.011, and a Signal-to-Noise Ratio of 22.90 dB, demonstrating high accuracy in shape reconstruction. Notably, this is the first study to leverage RGB-D data from actual spine surgeries, establishing a crucial baseline for future research in this domain. Shape completion holds significant promise for advancing computer- and robotic-assisted surgery by improving the precision of surgical robots and enhancing robotic perception in complex environments, especially with the integration of RGB-D cameras.

Looking ahead, while our method has the potential to drive advancements in machine perception and support both computer-assisted and robot-assisted surgeries, further steps are needed for clinical implementation. Hardware adjustments, real-time processing optimizations, and in-vivo validation will be crucial for transitioning this technology into patient care. Ensuring these advancements will be critical to successfully integrating these methods into real-world surgical settings. 

\section{Acknowledgements}
This work was supported by the FAROS project and has received funding from the European Union’s Horizon 2020 research and innovation program under grant agreement No. 101016985.

 \bibliographystyle{IEEEtran}
 \bibliography{mybibliography}

\begin{thebibliography}{10}
\providecommand{\url}[1]{#1}
\csname url@samestyle\endcsname
\providecommand{\newblock}{\relax}
\providecommand{\bibinfo}[2]{#2}
\providecommand{\BIBentrySTDinterwordspacing}{\spaceskip=0pt\relax}
\providecommand{\BIBentryALTinterwordstretchfactor}{4}
\providecommand{\BIBentryALTinterwordspacing}{\spaceskip=\fontdimen2\font plus
\BIBentryALTinterwordstretchfactor\fontdimen3\font minus \fontdimen4\font\relax}
\providecommand{\BIBforeignlanguage}[2]{{%
\expandafter\ifx\csname l@#1\endcsname\relax
\typeout{** WARNING: IEEEtran.bst: No hyphenation pattern has been}%
\typeout{** loaded for the language `#1'. Using the pattern for}%
\typeout{** the default language instead.}%
\else
\language=\csname l@#1\endcsname
\fi
#2}}
\providecommand{\BIBdecl}{\relax}
\BIBdecl

\bibitem{peng2020accuracy}
Y.-N. Peng, L.-C. Tsai, H.-C. Hsu, and C.-H. Kao, ``Accuracy of robot-assisted versus conventional freehand pedicle screw placement in spine surgery: a systematic review and meta-analysis of randomized controlled trials,'' \emph{Annals of translational medicine}, vol.~8, no.~13, 2020.

\bibitem{sielatycki2022state}
J.~A. Sielatycki, K.~Mitchell, E.~Leung, and R.~A. Lehman, ``State of the art review of new technologies in spine deformity surgery--robotics and navigation,'' \emph{Spine deformity}, pp. 1--13, 2022.

\bibitem{costa2016radiation}
F.~Costa, G.~Tosi, L.~Attuati, A.~Cardia, A.~Ortolina, M.~Grimaldi, F.~Galbusera, and M.~Fornari, ``Radiation exposure in spine surgery using an image-guided system based on intraoperative cone-beam computed tomography: analysis of 107 consecutive cases,'' \emph{Journal of Neurosurgery: Spine}, vol.~25, no.~5, pp. 654--659, 2016.

\bibitem{kendlbacher2022workflow}
P.~Kendlbacher, D.~Tkatschenko, M.~Czabanka, S.~Bayerl, G.~Bohner, J.~Woitzik, P.~Vajkoczy, and N.~Hecht, ``Workflow and performance of intraoperative ct, cone-beam ct, and robotic cone-beam ct for spinal navigation in 503 consecutive patients,'' \emph{Neurosurgical Focus}, vol.~52, no.~1, p.~E7, 2022.

\bibitem{li2023robot}
R.~Li, A.~Davoodi, Y.~Cai, K.~Niu, G.~Borghesan, N.~Cavalcanti, A.~Massalimova, F.~Carrillo, C.~J. Laux, M.~Farshad \emph{et~al.}, ``Robot-assisted ultrasound reconstruction for spine surgery: from bench-top to pre-clinical study,'' \emph{International journal of computer assisted radiology and surgery}, vol.~18, no.~9, pp. 1613--1623, 2023.

\bibitem{ji2015patient}
S.~Ji, X.~Fan, K.~D. Paulsen, D.~W. Roberts, S.~K. Mirza, and S.~S. Lollis, ``Patient registration using intraoperative stereovision in image-guided open spinal surgery,'' \emph{IEEE Transactions on Biomedical Engineering}, vol.~62, no.~9, pp. 2177--2186, 2015.

\bibitem{faraji2020machine}
Z.~Faraji-Dana, A.~L. Mariampillai, B.~A. Standish, V.~X. Yang, and M.~K. Leung, ``Machine-vision image-guided surgery for spinal and cranial procedures,'' in \emph{Handbook of robotic and image-guided surgery}.\hskip 1em plus 0.5em minus 0.4em\relax Elsevier, 2020, pp. 551--574.

\bibitem{liebmann2024automatic}
F.~Liebmann, M.~von Atzigen, D.~St{\"u}tz, J.~Wolf, L.~Zingg, D.~Suter, N.~A. Cavalcanti, L.~Leoty, H.~Esfandiari, J.~G. Snedeker \emph{et~al.}, ``Automatic registration with continuous pose updates for marker-less surgical navigation in spine surgery,'' \emph{Medical Image Analysis}, vol.~91, p. 103027, 2024.

\bibitem{fei2022comprehensive}
B.~Fei, W.~Yang, W.-M. Chen, Z.~Li, Y.~Li, T.~Ma, X.~Hu, and L.~Ma, ``Comprehensive review of deep learning-based 3d point cloud completion processing and analysis,'' \emph{IEEE Transactions on Intelligent Transportation Systems}, vol.~23, no.~12, pp. 22\,862--22\,883, 2022.

\bibitem{pan2021variational}
L.~Pan, X.~Chen, Z.~Cai, J.~Zhang, H.~Zhao, S.~Yi, and Z.~Liu, ``Variational relational point completion network,'' in \emph{Proceedings of the IEEE/CVF conference on computer vision and pattern recognition}, 2021, pp. 8524--8533.

\bibitem{yu2021pointr}
X.~Yu, Y.~Rao, Z.~Wang, Z.~Liu, J.~Lu, and J.~Zhou, ``Pointr: Diverse point cloud completion with geometry-aware transformers,'' in \emph{Proceedings of the IEEE/CVF international conference on computer vision}, 2021, pp. 12\,498--12\,507.

\bibitem{adapointr}
X.~Yu, Y.~Rao, Z.~Wang, J.~Lu, and J.~Zhou, ``Adapointr: Diverse point cloud completion with adaptive geometry-aware transformers,'' \emph{arXiv preprint arXiv:2301.04545}, 2023.

\bibitem{li2023anatomy}
J.~Li, A.~Pepe, G.~Luijten, C.~Schwarz-Gsaxner, J.~Kleesiek, and J.~Egger, ``Anatomy completor: A multi-class completion framework for 3d anatomy reconstruction,'' in \emph{International Workshop on Shape in Medical Imaging}.\hskip 1em plus 0.5em minus 0.4em\relax Springer, 2023, pp. 1--14.

\bibitem{beetz2023multi}
M.~Beetz, A.~Banerjee, J.~Ossenberg-Engels, and V.~Grau, ``Multi-class point cloud completion networks for 3d cardiac anatomy reconstruction from cine magnetic resonance images,'' \emph{Medical Image Analysis}, vol.~90, p. 102975, 2023.

\bibitem{gafencu2024shape}
M.-A. Gafencu, Y.~Velikova, M.~Saleh, T.~Ungi, N.~Navab, T.~Wendler, and M.~F. Azampour, ``Shape completion in the dark: completing vertebrae morphology from 3d ultrasound,'' \emph{International Journal of Computer Assisted Radiology and Surgery}, pp. 1--9, 2024.

\bibitem{liebmann2021spinedepth}
F.~Liebmann, D.~St{\"u}tz, D.~Suter, S.~Jecklin, J.~G. Snedeker, M.~Farshad, P.~F{\"u}rnstahl, and H.~Esfandiari, ``Spinedepth: A multi-modal data collection approach for automatic labelling and intraoperative spinal shape reconstruction based on rgb-d data,'' \emph{Journal of Imaging}, vol.~7, no.~9, p. 164, 2021.

\bibitem{yolov8_ultralytics}
\BIBentryALTinterwordspacing
G.~Jocher, A.~Chaurasia, and J.~Qiu, ``Ultralytics yolov8,'' \emph{Ultralytics YOLOv8}, 2023. [Online]. Available: \url{https://github.com/ultralytics/ultralytics}
\BIBentrySTDinterwordspacing

\bibitem{mendelsohn2016patient}
D.~Mendelsohn, J.~Strelzow, N.~Dea, N.~L. Ford, J.~Batke, A.~Pennington, K.~Yang, T.~Ailon, M.~Boyd, M.~Dvorak \emph{et~al.}, ``Patient and surgeon radiation exposure during spinal instrumentation using intraoperative computed tomography-based navigation,'' \emph{The Spine Journal}, vol.~16, no.~3, pp. 343--354, 2016.

\bibitem{kirillov2023segany}
A.~Kirillov, E.~Mintun, N.~Ravi, H.~Mao, C.~Rolland, L.~Gustafson, T.~Xiao, S.~Whitehead, A.~C. Berg, W.-Y. Lo, P.~Doll{\'a}r, and R.~Girshick, ``Segment anything,'' \emph{arXiv:2304.02643}, 2023.

\bibitem{qi2017pointnet++}
C.~R. Qi, L.~Yi, H.~Su, and L.~J. Guibas, ``Pointnet++: Deep hierarchical feature learning on point sets in a metric space,'' \emph{Advances in neural information processing systems}, vol.~30, 2017.

\bibitem{qi2017pointnet}
C.~R. Qi, H.~Su, K.~Mo, and L.~J. Guibas, ``Pointnet: Deep learning on point sets for 3d classification and segmentation,'' in \emph{Proceedings of the IEEE conference on computer vision and pattern recognition}, 2017, pp. 652--660.

\bibitem{open3d}
Q.-Y. Zhou, J.~Park, and V.~Koltun, ``{Open3D}: {A} modern library for {3D} data processing,'' \emph{arXiv:1801.09847}, 2018.

\bibitem{wang2019dynamic}
Y.~Wang, Y.~Sun, Z.~Liu, S.~E. Sarma, M.~M. Bronstein, and J.~M. Solomon, ``Dynamic graph cnn for learning on point clouds,'' \emph{ACM Transactions on Graphics (tog)}, vol.~38, no.~5, pp. 1--12, 2019.

\bibitem{tatarchenko2019single}
M.~Tatarchenko, S.~R. Richter, R.~Ranftl, Z.~Li, V.~Koltun, and T.~Brox, ``What do single-view 3d reconstruction networks learn?'' in \emph{Proceedings of the IEEE/CVF conference on computer vision and pattern recognition}, 2019, pp. 3405--3414.

\bibitem{zeng20193d}
J.~Zeng, G.~Cheung, M.~Ng, J.~Pang, and C.~Yang, ``3d point cloud denoising using graph laplacian regularization of a low dimensional manifold model,'' \emph{IEEE Transactions on Image Processing}, vol.~29, pp. 3474--3489, 2019.

\bibitem{kazhdan2006poisson}
M.~Kazhdan, M.~Bolitho, and H.~Hoppe, ``Poisson surface reconstruction,'' in \emph{Proceedings of the fourth Eurographics symposium on Geometry processing}, vol.~7, no.~4, 2006.

\bibitem{meng2020learning}
D.~Meng, M.~Keller, E.~Boyer, M.~Black, and S.~Pujades, ``Learning a statistical full spine model from partial observations,'' in \emph{Shape in Medical Imaging: International Workshop, ShapeMI 2020, Held in Conjunction with MICCAI 2020, Lima, Peru, October 4, 2020, Proceedings}.\hskip 1em plus 0.5em minus 0.4em\relax Springer, 2020, pp. 122--133.

\bibitem{caprara2022bone}
S.~Caprara, M.-R. Fasser, J.~M. Spirig, J.~Widmer, J.~G. Snedeker, M.~Farshad, and M.~Senteler, ``Bone density optimized pedicle screw instrumentation improves screw pull-out force in lumbar vertebrae,'' \emph{Computer Methods in Biomechanics and Biomedical Engineering}, vol.~25, no.~4, pp. 464--474, 2022.

\bibitem{yanni2021real}
D.~S. Yanni, B.~M. Ozgur, R.~G. Louis, Y.~Shekhtman, R.~R. Iyer, V.~Boddapati, A.~Iyer, P.~D. Patel, R.~Jani, M.~Cummock \emph{et~al.}, ``Real-time navigation guidance with intraoperative ct imaging for pedicle screw placement using an augmented reality head-mounted display: A proof-of-concept study,'' \emph{Neurosurgical Focus}, vol.~51, no.~2, p. E11, 2021.

\end{thebibliography}

\end{document}